\documentclass[9pt,twocolumn,twoside]{pnas-new}
\templatetype{pnasresearcharticle} % Choose template 
\title{Dynamics of Droplet Generation from Corneal Tear Film during Non-contact Eye Procedure in the Context of COVID-19}

\author[a]{Durbar Roy}
\author[a]{Sophia M} 
\author[a]{Abdur Rasheed}
\author[b]{Prasenjit Kabi}
\author[c]{Abhijit Sinha Roy}
\author[d]{Rohit Shetty}
\author[a,1]{Saptarshi Basu}

\affil[a]{Department of Mechanical Engineering, Indian Institute of Science, Bengaluru, 560012, India}
\affil[b]{Interdisciplinary Centre for Energy Research, Indian Institute of Science, Bengaluru, 560012, India}
\affil[c]{Imaging, Biomechanics and Mathematical Modelling Solutions lab, Narayana Nethralaya Foundation, Bangalore, India}
\affil[d]{Narayana Nethralaya Eye Hospital, Bangalore, India}
\leadauthor{Roy} 

\correspondingauthor{\textsuperscript{1}To whom correspondence should be addressed. E-mail: sbasu@iisc.ac.in}


\begin{abstract}
Non-invasive medical diagnostics demonstrate a propensity for droplet generation and should be studied to devise risk mitigation strategies against the spread of the SARS-CoV-2 virus.
We investigate the air-puff tonometry, which uses a short-timed air-puff to applanate the human eye in a bid to detect the early onset of glaucoma by measuring the intraocular pressure. The air-puff consists of a vortex trailed by a high-speed jet. High-speed imaging of the eye during a typical tonometry measurement reveals a sequence of events starting with the interaction between the tear layer and the air puff leading to an initial sheet ejection. It is immediately followed by the trailing jet applanating the central
corneal section, causing capillary waves to form and interact with the highly 3D transient expanding
sheet. Such interaction with the capillary waves and the surrounding airfield due to the trailing jet causes the expanding sheet to undergo bag breakup, finger formation by
Rayleigh Taylor instability and further break up into subsequent droplets by Rayleigh Plateau
instability, which eventually splashes onto nearby objects, potentially forming fomites or aerosols which can lead to infections. The complex spatiotemporal phenomenon is carefully documented by rigorous experiments and
corroborated using comprehensive theoretical analysis.
\end{abstract} 

\begin{document}

\maketitle
\ifthenelse{\boolean{shortarticle}}{\ifthenelse{\boolean{singlecolumn}}{\abscontentformatted}{\abscontent}}{}

\begin{SCfigure*}
\centering
\includegraphics[scale=1]{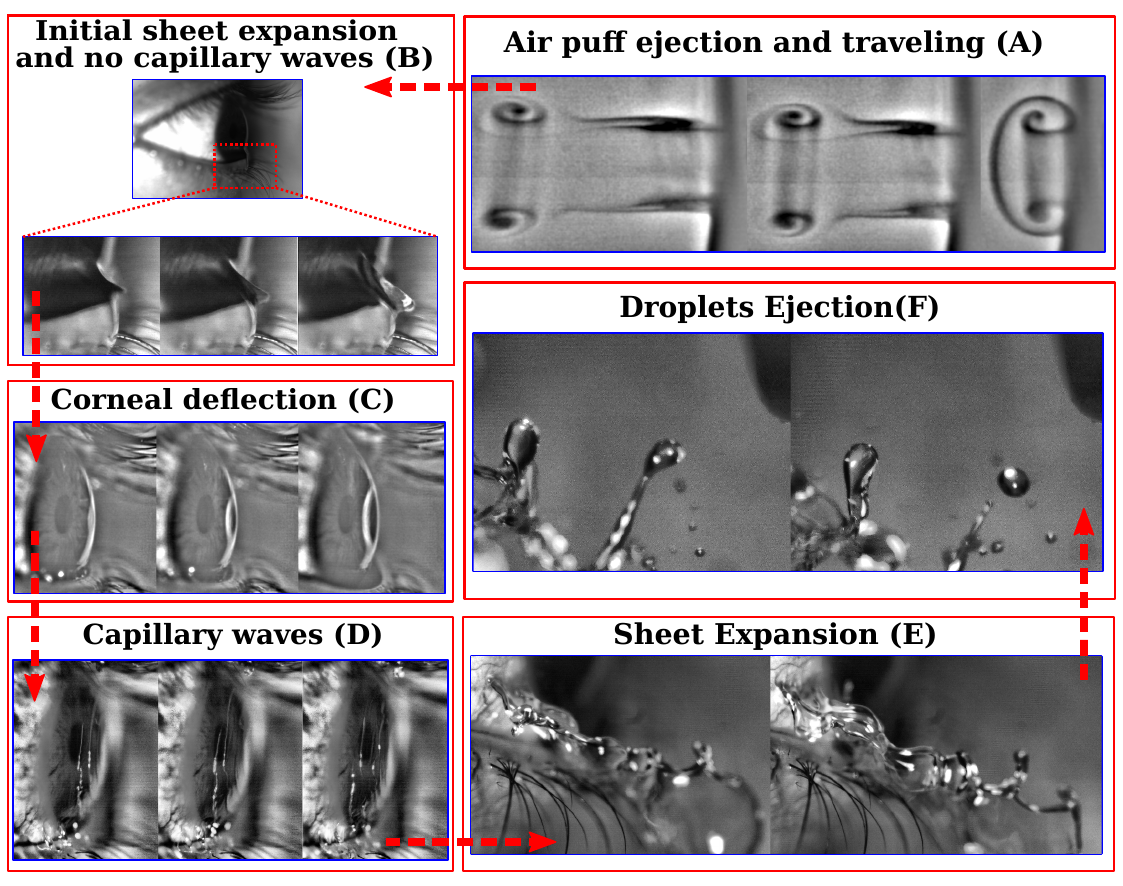}
\caption{Different Phases involved during the non-contact tonometry from a fluid dynamical perspective has been shown. Phase A represents the air puff generation from the nozzle and its propagation. Phase B consists of the initial sheet expansion due to sudden pressure reduction due to the approaching leading vortex. Phase C denotes the central corneal deflection due to the trailing jet. Phase D shows the detection and propagation of capillary waves on the tear film surface. Phase E shows the later sheet expansion, and Phase F shows the droplet breakup regime due to Rayleigh Plateau and its propagation under the influence of its linear momentum, gravity, and air resistance.}
\label{Fig. 1.}
\end{SCfigure*}
% If your first paragraph (i.e. with the \dropcap) contains a list environment (quote, quotation, theorem, definition, enumerate, itemize...), the line after the list may have some extra indentation. If this is the case, add \parshape=0 to the end of the list environment.

%\section*{Introduction}
\dropcap{C}ovid-19 is the current global health pandemic. Mostly the virus SARS-CoV-2 proliferates through droplets of saliva and discharge of the nose. It happens through coughing, talking, breathing and sneezing of a COVID-infected person. Many studies related to coughs and sneezing in both the clinical/medical community and the scientific/engineering community like the works of Scharfman et al. , Bourouiba et al., to name a few
%\cite{scharfman2016visualization}\cite{bourouiba2014violent}.
\cite{scharfman2016visualization, bourouiba2014violent} studied the mechanisms involved.
Scharfman et al., for the first time, showed that the breakup of droplets happens outside the respiratory tracts in violent exhalation processes. They further showed various kinds of droplet breakup mechanisms like sheet breakup, bag breakup, and ligament breakup during such processes.
Bourouiba et al. also studied the fluid dynamics of violent expiratory events like coughing and sneezing using high-speed scattering imaging techniques. They discovered that the flow is highly turbulent and multiphase. Further, a theoretical analysis of pathogen bearing droplets interacting with the turbulent momentum puff was conducted. The effect of respiratory droplets on the spread of Covid-19 was modelled by Chaudhari et al \cite{chaudhuri2020modeling}.
The virus transmission can be minimized by adequately understanding and studying the various causes of droplet generation from an infected individual and finding solutions to protect everyone from such droplets. 
This work deals with a non-evasive medical procedure called non-contact tonometry.
Tonometry is a technique used by ophthalmologists to measure intraocular pressure that gives a good idea of whether any patient is suffering from glaucoma or not. Further intraocular pressure also helps to diagnose various other eye-related conditions that may be helpful to the specialist\cite{brandt2007central}. The first most accurate IOP measuring device was the Maklakoff applanation tonometer\cite{draeger1967principle}. These early tonometers were the contact type tonometer. Medical practitioners and researchers modified and developed the tonometer through several generations increasing the reliability and accuracy of the device.  
 The new breed of tonometers like McKay-Marg tonometer, Tono-Pen, pneumatonometer, and air puff tonometer to name a few have gained popularity among medical practitioners \cite{stamper2011history}. This work uses an air-puff type tonometer, which is a non-contact type. Many studies have been conducted on the working principle and the development of non-contact tonometers. Ophthalmologists have conducted many clinical studies about IOP, its variation with time, its effect on different eye-related medical conditions like glaucoma, corneal stiffness, and thickness, to mention a few. Some work related to the corneal deflection in response to an air puff tonometry has also been studied using experimental and computational techniques\cite{simonini2016influence}. The complicated structure of the human eye makes theoretical modeling quite challenging\cite{delmonte2011anatomy}.
Scientists have conducted preliminary clinical studies decades ago and showed that droplets maybe generated during non-contact tonometry\cite{britt1991microaerosol}. Some clinical studies shows that the SARS-CoV-2 virus can be present in human tears like the work of Sadhu et al.\cite{sadhu2020covid}. Such aerosols, may lead to the transmission of the virus undetected.
 The literature on the fluid mechanics of tonometry is quite sparse. The spatiotemporal scales involved are quite small, and hence resolving the phenomena is quite challenging. The air puff emanating of a non contact tonometer is essentially a vortex ring. There have been some studies on vortex ring impact on thin liquid films
\cite{gendrich1997whole}, but the resulting fluid mechanics due to vortex ring interaction with a human eye tear film has never been probed in detail. Previously we have carried out some preliminary studies related to non-contact tonometry  \cite{shetty2020quantitative}.
This work provides some rigorous comprehensive insights into non-contact tonometry using experimental and theoretical investigations performed on real human subjects. High-speed imaging and visualizations were carried out to decipher the underlying fluid dynamics in sufficient details for dry and watery eye conditions.
Fig. 1 provides an overview of the entire time series of events during non-contact tonometry.
Different phases have been identified on the basis of the order of events (according to temporal history.)
The phenomenon starts with an air puff (a leading vortex with a trailing jet) ejected from the nozzle head of the tonometer. The leading vortex travels towards the eye, increasing the local air velocity field adjacent to the eye, which reduces the local pressure outside the tear layer. The transient pressure reduction causes the tear film accumulated at the bottom portion of the eye to be ejected as a sheet especially in the case of watery eye. Meanwhile, the trailing jet impacts the cornea causing a detectable deformation of the eye. The deformation leads to detectable capillary waves on the eye surface, which travel outwards in the form of concentric rings \cite{ceniceros1999dynamic}. The initial sheet is reinforced by the capillary waves, expand and undergoes bag breakup \cite{jain2015secondary} and Rayleigh Taylor instabilities \cite{bussmann2000modeling,drazin1981hydrodynamic,drazin2002introduction,chandrasekhar2013hydrodynamic}. The Rayleigh Taylor instability causes  finger-like structure to form and grow, which further undergoes Rayleigh Plateau breakup \cite{drazin1981hydrodynamic,drazin2002introduction,chandrasekhar2013hydrodynamic} into droplets. The bag breakup and Rayleigh Plateau breakup produces droplets of various sizes and velocity. These droplets can then travel in various directions under the effect of its linear momentum, acceleration due to gravity, and air resistance. The details of each phase have been described in the results and discussion sections and also illustrated in Fig. 1. 
The main aim of this work is on the fluid dynamics rather than on the corneal deflection since a lot of literature is already available \cite{simonini2016influence,delmonte2011anatomy}. We study the interaction between the air puff and the human eye and the associated fluid dynamics at different Spatio-temporal scales in sufficient detail. Further, to assess the spread of the COVID-19 virus, the size distribution and the velocity distribution of the droplets that are generated during the non-contact tonometry procedure are reported, which may help the opthalmologist and eye care practitioners to follow safety protocols while undertaking these procedures. 

\begin{figure}%[tbhp]
\centering
\includegraphics[width=8.7cm]{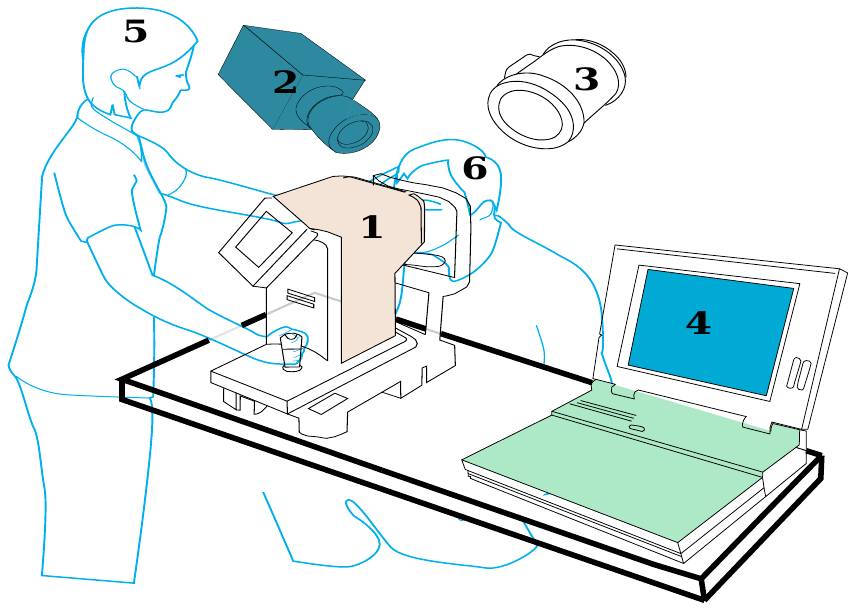}
\caption{Schematic of the experimental setup. The different components of the setup are 1: Non Contact Tonometer, 2: High Speed Camera, 3: Light Source, 4: Computer, 5: Tonometer Operator, 6: Human Subject}
\label{Fig. 2.}
\end{figure}

\section*{Materials and Methods}
The schematic of the experimental setup is shown in Fig. 2. The experiments were conducted on four different human subjects of various age groups and gender (one female and three males). High-speed images were acquired using Photron Mini UX100 and a high-intensity light source Veritas Mini Constellation at 2000 frames per second with a spatial resolution of 512 pixels $\times$ 320 pixels. The images were taken from different views like side view, orthographic views, oblique views, inclined views, and back views to capture as much information as possible since the fluid dynamics involved are highly transient and three-dimensional in nature.    
The tonometer's nozzle was kept at a working distance of approximately $11mm$ from the human eye, which is the standard protocol for NCT200 to give a correct reading of intraocular pressure. The high-intensity light source was placed in different strategic locations to give the desired lighting for different views. The initial vortex and jet were observed by smoke flow visualization and scattering techniques. The trailing jet was also visualized using alumina particles at various frames per second using scattering techniques.
However, a dead goat eye was placed in the place of the real human subject for smoke flow visualization in order to characterize the air puff coming out of the tonometer and to understand the roles of the leading vortex and the trailing jet. Experiments were conducted on two different eye conditions: dry and wet. Dry eye conditions were studied so that we have a comprehensive understanding of the corneal deflection and observe whether any droplets are ejected in dry conditions. The wet eye conditions were simulated using eye drops that were prescribed by an ophthalmologist.
Some industrial brands of eye drops that were used are Refresh Tears, Lubrex, Trehalube, Systrane Ultra.
The drops were used on the human subjects just before the tonometry measurement to mimic watery eye condition.
The images were acquired with the Photron FastCam Viewer (software package 4.0.3.4.) and the raw images were processed using a combination of open-source tools, ImageJ \cite{schneider2012nih} and python programming language \cite{10.5555/1593511}. Preprocessing of the raw images were performed using FFT Bandpass Filter \cite{baker1980multiple} and by using CLAHE (Contrast Limited Adaptive Histogram Equalization) \cite{setiawan2013color}.
The significant structures were filtered down to 40 pixels, and small structures up to 3 pixels were used for the FFT Bandpass Filter.
For the CLAHE, the block size used was 127 pixels, histogram bins of 256, and a maximum slope of 3 were used. The various kinematic parameters involved were then tracked, and the results were post-processed using Python programming language\cite{10.5555/1593511}. The raw data of the kinematic parameters were converted into probability density function using kernel density estimation.
%A SI video is provided for a complete overview of all the processes. The data files are provided as csv files in SI datasets.

\section*{Results and Discussions}
Fig. 1. demarcates the various phases of the tonometry process from a fluid dynamical perspective based on timescale.
%(See SI Video for a complete overview)
Phase A consists of the leading vortex approaching the human eye, and covering a working distance of approximately $11mm$ in approximately $2.6ms$. Phase B starts with the initial sheet ejection and ends till the the effect of capillary waves is felt on the initial sheet. This phase lasts for approximately $3ms$. Phase C begins with the corneal deflection due to the trailing jet impinging the cornea and this phase lasts up to $7.5ms$. Phase D starts when the first detectable capillary waves are seen on the surface of the eye and this phase lasts up to $3ms$. Phase E consists of all the interesting fluid mechanical phenomena related to sheet expansion, Rayleigh Taylor waves, bag formation, finger like structure formation. This phase lasts for approximately $5ms$. The last phase F consists of the disintegration into the droplets due to Rayleigh plateau breakup. This phase lasts for approximately $3.5ms$. This phase and beyond deals with the droplets of various size and velocity propagating under the combined effect of the initial linear momentum of the droplet, the acceleration due to gravity and effects of air-resistance. All the phases are discussed below in their respective subsections.
%\begin{figure}
%    \centering
%    \includegraphics[scale=0.30]{./TIMELINE_SCHEMATICS_FINAL.pdf}
%    \caption{Schematics representation of different phases}
    %\label{fig:my_label}
%\end{figure}

\subsection*{Vortex ring approaches the eye (Phase A)}
The non-contact tonometry procedure starts with the ejection of an air puff, which is essentially a leading vortex followed by a trailing jet from the tonometer nozzle. The position of the leading vortex was tracked from the smoke flow visualization images. The leading vortex propagates before it interacts with the tear film of the human eye.
Fig. 3 depicts the air puff characteristics. Fig. 3(a) shows the smoke flow visualization of the leading propagating vortex and the trailing jet.
The vortex propagation was tracked by measuring the horizontal position $x$ of the vortex core (shown as a green dot), from a predefined reference line (shown as a red dotted line) which is the exit plane of the tonometer nozzle, as a function of time. Two other parameters, $r_{1}$ and $r_{2}$ were also tracked to understand the vortex's radial dynamics. $r_{1}$ gives an estimate of the vortex's outer envelope, and $r_{2}$ gives us an idea of the internal length scale of the propagating vortex.
Fig. 4(a) shows the vortex's propagation characteristics as it approaches an obstacle, which was essentially a goat's eye. Fig. 4(b) shows the temporal evolution of $r_{1}$ and $r_{2}$. $r_{1}$ and $r_{2}$ increases linearly with time, showing a small expansion of the leading vortex. Both $r_{1}$ and $r_{2}$ are increasing at the same rate since the slope is approximately identical, as can be observed from Fig. 4(b). The sudden jumps at approximately $2ms$ in both Fig. 4(a) and 4(b) denote the time at which the leading vortex starts interacting with the eye. The leading vortex takes approximately $3ms$ to cover a working distance of approximately $11mm$ for the human subjects. The leading vortex average velocity scale is approximately $V_{vortex}{\sim}5m/s$.
\begin{figure}
    \centering
    \includegraphics[scale=1]{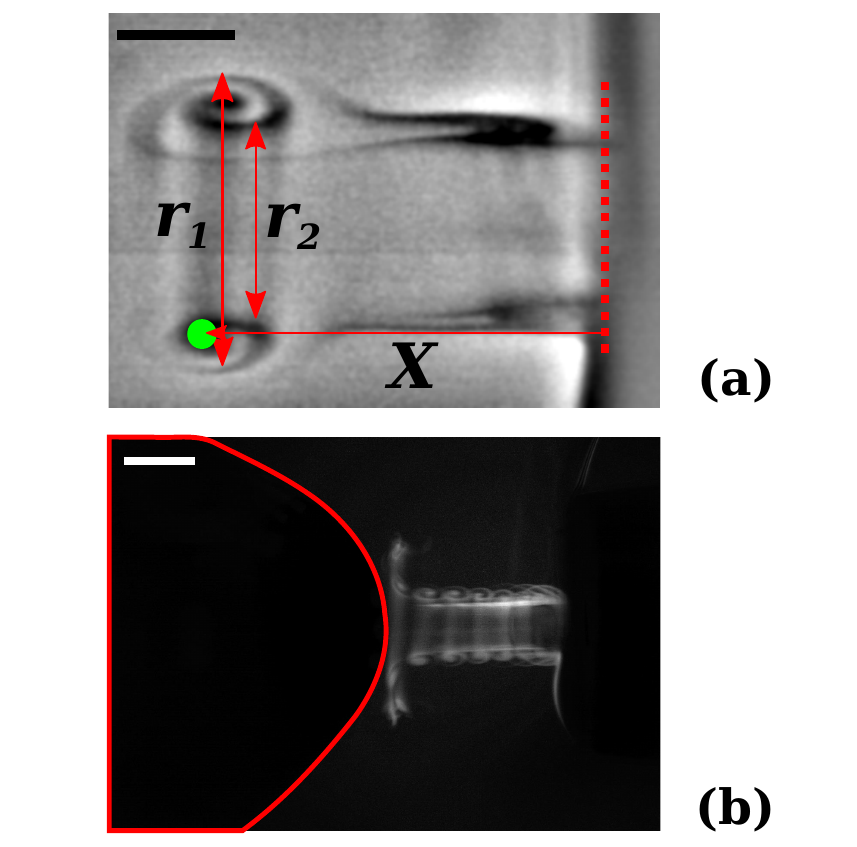}
    \caption{Air puff characterization. (a) shows the smoke flow visualization of the leading vortex and the trailing jet along with the tracking parameters. The scale bar denotes 3mm in actual length, (b) depicts the propagation of the leading vortex towards a goat eye. The scale bar denotes 6mm in actual length.}
    \label{Fig. 3.}
\end{figure}
\begin{figure}
    \centering
    \includegraphics[scale=1]{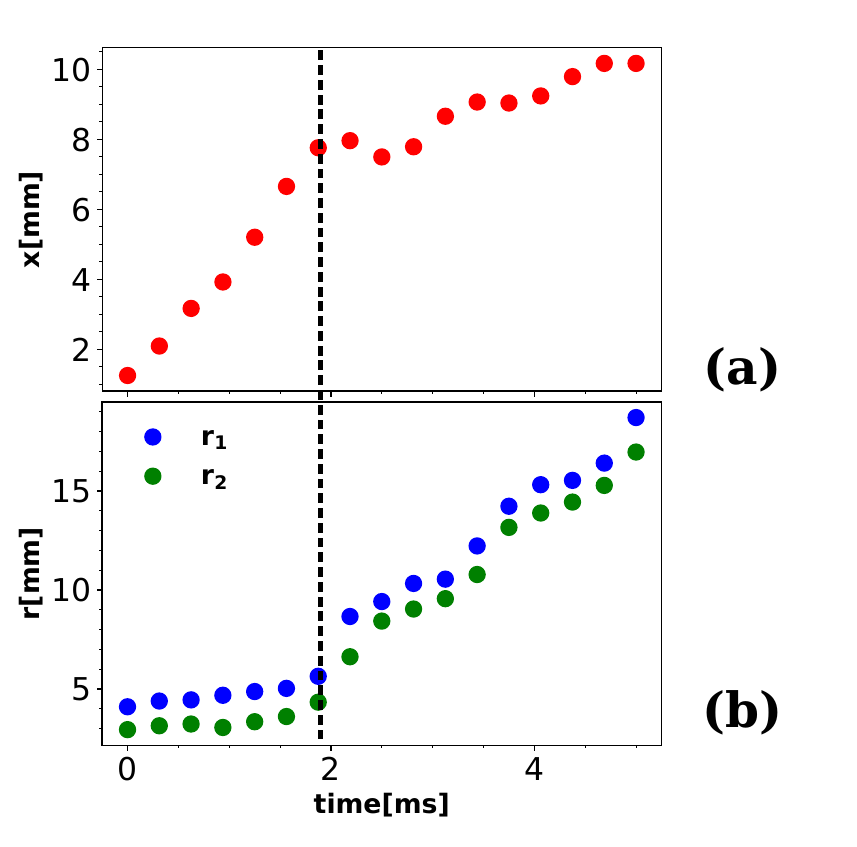}
    \caption{Air puff characterization. (a) displacement of the leading vortex vs time in ms, (b) radial expansion of the leading vortex vs time in ms, showing both outer diameter and inner diameter of the leading vortex.}
    \label{Fig. 4.}
\end{figure}
\subsection*{Initial sheet ejection (Phase B)}
\begin{figure}
    \centering
    \includegraphics[scale=0.75]{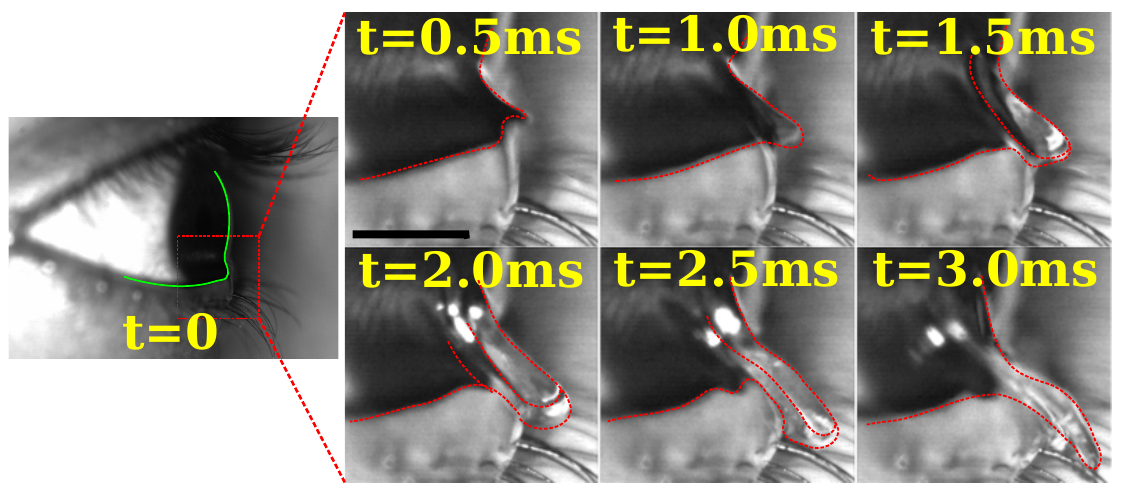}
    \caption{Initial sheet expansion and its temporal evolution shown as a series of snapshots 0.5ms apart. The scale bar denotes 3mm.}
    \label{Fig. 5.}
\end{figure}
        As the leading vortex approaches the human eye, the velocity field locally adjacent to the eye increases with time, which causes a sudden reduction in the local pressure.
        The effect of this reduced pressure is different for dry and wet eye conditions. For dry eye, the tear distribution is uniform across the entire corneal surface, and the interaction adhesive force between the tear film and the cornea is quite strong. Hence no sheet-like structures are seen in the dry eye condition, and hence no droplets are ejected out. However, for a teary eye condition as simulated using eye drops for the experiments, the excess tear is located in the lower part of the cornea. Hence the tear film thickness is not uniform, as was the case for dry eye. The tear film has a distribution that is thicker at the bottom. Due to the higher thickness of the film, the adhesive force between the cornea and the film is smaller compared to the dry eye conditions.
        Hence the reduced outer field pressure causes fluid sheet ejection in the case of a wet eye.
        Fig. 5 shows the temporal evolution of the sheet, along with the corresponding time scale.
        The sheet's initial sheet ejection velocity can be modeled using an in-viscid analysis because the effect of viscosity can be neglected. The theoretical analysis of the sheet expansion is provided in Phase E.
\subsection*{Corneal deflection (Phase C)}
The leading vortex is followed by a trailing jet. The trailing jet upon impact causes a central corneal deflection.
Fig. 6 shows the entire image sequence of the corneal deflection process. The entire deformation process has an overall time scale of approximately $7.5ms$. Fig. 7 shows the central corneal deflection curve as a function of time. The corneal deflection increases to a maximum and then returns to the undeformed state. The curve peaks at approximately $4ms$, and the maximum corneal deflection varies from $0.6mm$ to $0.9mm$. The scatter present in the displacement curve is due to the different intraocular pressure of the eye. The intraocular pressure is like the blood pressure of an individual, which is biological in origin and is dependent on many factors and varies within a band for an individual. Some of the factors are the time of the day, conditions before and after drinking water, to name a few \cite{buddle2014day}. To exactly model the corneal deflection, the anatomy of the cornea has to be understood adequately \cite{delmonte2011anatomy}. Some literature is available to model the deflection of the cornea in extensive detail \cite{simonini2016influence}. This work focuses on the corneal deflection qualitatively only.  
\begin{figure}
    \centering
    \includegraphics[scale=1]{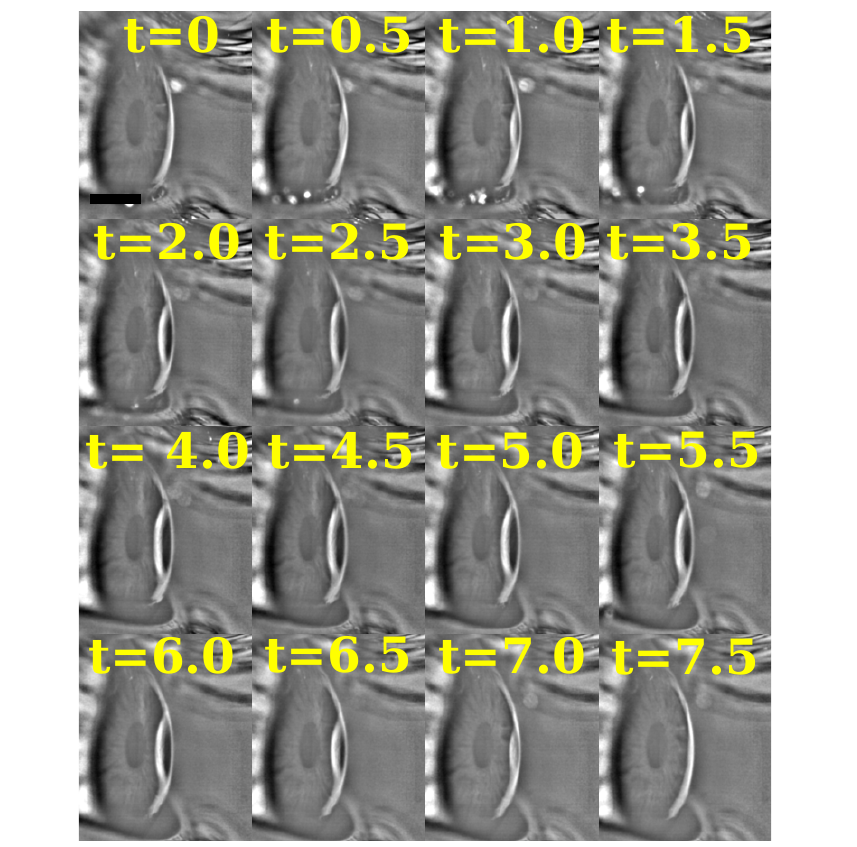}
    \caption{Closeup high speed snapshots of the entire corneal deflection process. All time are in milliseconds (ms). The scale bar denotes 3mm.}
    \label{Fig. 6.}
\end{figure}

\begin{figure}
    \centering
    \includegraphics[scale=1]{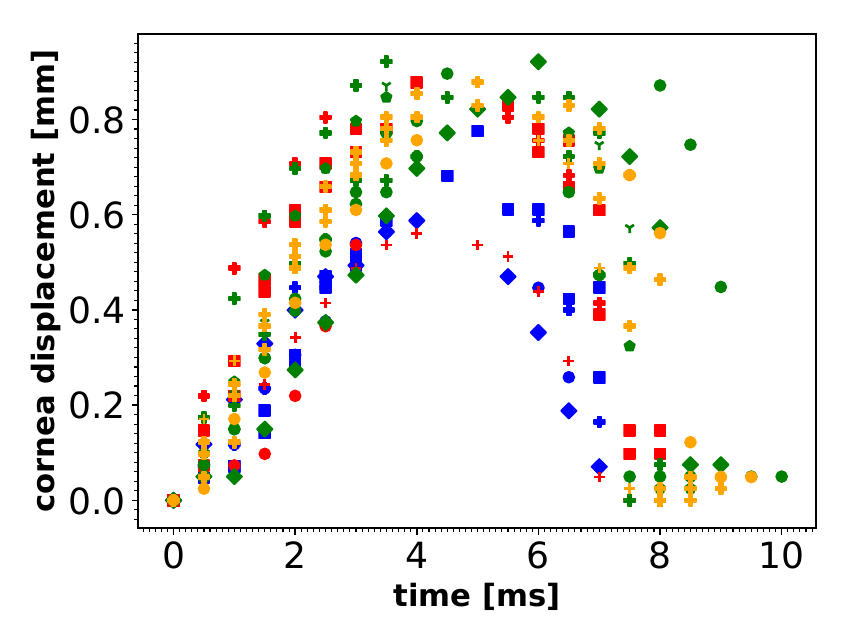}
    \caption{Corneal deflection curve in mm as a function of time in ms.} 
    \label{Fig. 7.}
\end{figure}

\subsection*{Capillary waves (Phase D)}
The impact of the trailing jet causes a central corneal deflection along with the detectable capillary waves. Capillary waves are caused due to pressure variation on the top surface of the tear film. The pressure variations form ripples on the free surface of the liquid, which travels outwards. These ripples produce film thickness variation of the corneal tear film. The ripples, also known as capillary waves, are an outcome of the interplay between the external pressure (due to the trailing jet and its associated airflow field) and the air-water surface tension, which acts as a restoring force. The waves/ripples generated on the tear film are approximately symmetrical about the central spot where the trailing jet impacted the cornea. A radial symmetry exists at least on a local small Spatio-temporal scale, and hence a one-dimensional model is a good enough approximation to decipher the relevant velocity scales involved.
The dispersion relations for a pure capillary waves (1-D relation)is given by
  \begin{equation}
    {\omega}=\sqrt{\frac{\sigma_{aw}}{{\rho}_a+{\rho}_w}}k^{3/2}
  \end{equation}
  Where ${\omega}$ is the growth rate of a disturbance, ${\sigma}_{aw}$ is the air-water surface tension, ${\rho_a}$ is the density of air, ${\rho_w}$ is the density of water, ${k}=\frac{2{\pi}}{\lambda}$ is the wavenumber which is inversely proportional to the wavelength of the capillary waves. A pure capillary wave's growth rate is directly proportional to the wavenumber raised to three-half powers. This implies that smaller wavelengths have higher growth rates compared to larger ones. The growth rate also depends on the material properties. ${\omega}$ is directly proportional to the square root of the air-water surface tension. This implies that higher surface tension liquids will have higher growth rates. ${\omega}$ is also inversely proportional to the square root of the sum of air density and water density.
  The phase velocity is the velocity with which capillary waves of a given wavenumber/wavelength propagates and is given as the ratio of the growth rate of the disturbance to the wavenumber of the ripples.
  \begin{equation}
v_{phase}=\frac{\omega}{k}=\sqrt{\frac{\sigma_{aw}}{{\rho}_a+{\rho}_w}}k^{1/2}
\end{equation}
and the group velocity is given as the first derivative of the growth rate with respect to wavenumber.
\begin{equation}
v_{group}=\frac{d\omega}{dk}=\frac{3}{2}\sqrt{\frac{\sigma_{aw}}{{\rho}_a+{\rho}_w}}k^{1/2}
\end{equation}
  The group velocity helps to calculate the velocity of an envelope which contains waves of various wavelengths.
  Equation [2] and [3] have the same variation with wavenumber and the dependence is square root of wavenumber. This implies that short wavelengths have higher velocity of propagation compared to long wavelengths.
\begin{figure}
    \centering
    \includegraphics[scale=1]{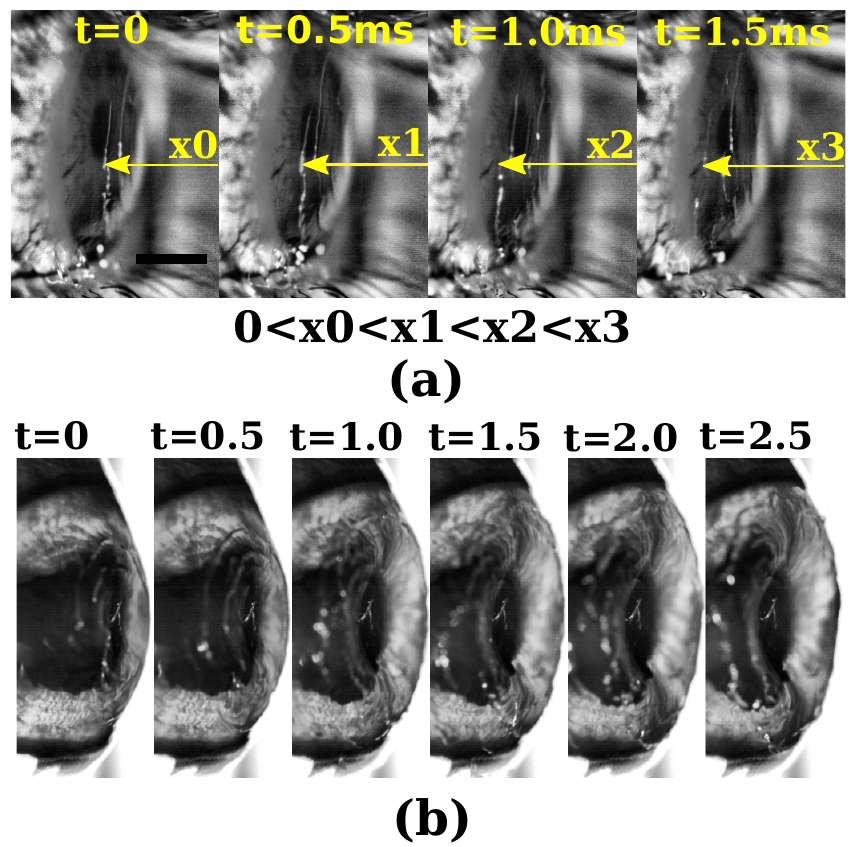}
    \caption{(a) Capillary waves travelling on the surface of a human eye. The scale bar denotes 3mm. (b) Capillary waves seen on a cadaver goat eye.}
    \label{Fig. 8.}
\end{figure}

\begin{figure}
    \centering
    \includegraphics[scale=1]{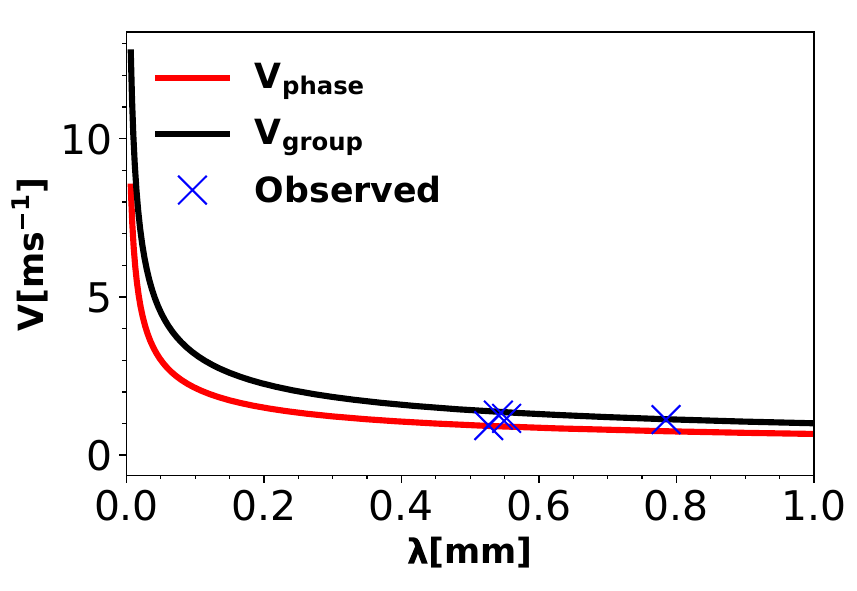}
    \caption{Theoretical phase velocity and group velocity as a function of wavelength along with the experimental velocities.}
    \label{Fig. 9.}
\end{figure}

Fig. 8(a) depicts the timeline of the propagation of the capillary waves. The peak of a capillary wave was tracked from a fixed reference datum and the average velocity of propagation was calculated using the forward difference approximation of the first derivative of position from displacement time relation. We did some experiments to verify the generality of the idea of capillary waves on cadaver goat eye. The time sequence of the resulting capillary waves on the goat eye is shown in Fig. 8(b). The phase velocities and group velocities are plotted as a function of wavelength as depicted in Fig. 9, respectively, using equation [2] and [3], respectively. 
The experimentally measured values are shown in Fig. 9, which are of the same order of magnitude as that obtained theoretically. The order of magnitude is approx $V{\sim}1.1m/s$, which corroborates well with the theoretical framework.

\subsection*{Expansion of the sheet/ RT Waves/ Bag formation (Phase E)}
\subsubsection*{Dynamics of Sheet Expansion}
As the leading vortex approaches the eye, the local velocity field just outside of the tear film increases.
This leads to a very thin boundary layer during the first $2ms$, which is of the order of approximately $0.0242mm$. Outside this length scale, viscous effects can be neglected, and hence the sheet expansion can be modeled using inviscid theory. 
        Hence using Eulers equation we have:
        \begin{equation}
  {\rho}_{w}\frac{{d}V_{sheet}}{dt}=-{\frac{{\partial}p}{{\partial}r}}
\end{equation}
where ${\rho}_w$ denotes the density of water, ${V_{sheet}}$ is the sheet ejection velocity, $dt$ denotes the elementary timescale, $p$ denotes pressure and $r$ is an characteristic length scale characterizing the outer envelope of the sheet. Figure 10(a) shows the schematic of the sheet expansion. $p_{in}$ is the initial pressure inside the tear film at static condition. The initial tear film thickness is $r(t=0)=F_{0}$.
Applying scaling analysis on the Eulers equation we have
\begin{equation}
\frac{{d}V_{sheet}}{{d}t}{\sim}{\frac{1}{{\rho}_w}}\frac{p_{in}-p_{out}}{\Delta{r}}
\end{equation}
Assuming $p_{in}{\simeq}p^{*}$, which is the initial pressure inside the static tear layer before the interaction with the air-puff. A pressure difference exists across the interface due to air-water surface tension and the tear film's curvature at the lower part of the eye in the static condition. However, due to an increasing outer velocity field in the air medium, the outer pressure $p_{out}$ in the airfield reduces by an amount given by the Bernoulli's equation kinetic energy per unit volume term
\begin{equation}
p_{out}{\simeq}p^{*}-\frac{{\sigma}_{aw}}{r}-\frac{1}{2}{\rho}_{air}V_{air}^2
\end{equation}
\begin{equation}
p_{in}-p_{out}{\simeq}{\frac{{\sigma}_{aw}}{r}}+\frac{1}{2}{\rho}_{air}V_{air}^2
\end{equation}
Substituting equation [7] in equation [5] we have,
%$$
%{{\Delta}V_{sheet}}=\left(\frac{{\sigma}_{aw}{\Delta}t}{{\rho}_{w}r^2}+\frac{1}{2}\frac{{\rho}_{air}}{{\rho}_{w}}\frac{V_{air}^2{\Delta}t}{r}\right)
%$$

\begin{equation}
\frac{d{V}_{sheet}}{dt}=\frac{{\sigma}_{aw}}{{\rho}_{w}r^2}+\frac{1}{2}\frac{{\rho}_a}{{\rho}_w}\frac{{V}_{air}^2}{r}
\end{equation}
Equation [8] depicts the sheet's acceleration and is essentially a form of Newtons' second law
of motion. This equation is one of the central results of the current work. The first term on the right-hand side is the acceleration due to surface tension and the curvatur of the interface. The second term is the acceleration due to the air velocity field. The sheet velocity is given by the first derivative of $r$,  
$
V_{sheet}=\frac{dr}{dt}
$.
Substituting the value of $V_{sheet}$ in terms of first derivative of $r$, we have

\begin{equation}
\frac{d^2r}{dt^2}=\frac{a}{2r^2}+\frac{b}{2r}
\end{equation}
The above equation has been converted into two first order equations and are solved numerically. The first order derivatives were discretized using backward euler method.
\begin{equation}
\dot{X_1}=X_2
\end{equation}

\begin{equation}
\dot{X_2}=\frac{a}{2X_1^2}+\frac{b}{2X_1}
\end{equation}
The initial conditions are the position initial condition $X_1(t=0)$ and the velocity initial condition $X_2(t=0)$.

The position initial condition is given as
\begin{equation}
X_1(t=0)=F_0
\end{equation}
where $F_0$ denotes the initial film thickness.

The velocity initial condition is given by
\begin{equation}
X_2(t=0)=0
\end{equation}

where
$
a=\frac{2{\sigma}_{aw}}{{\rho}_w}
$
 which is a constant and is twice the ratio of air-water surface tension to the density of water
,
$
b=\frac{{\rho}_a}{{\rho}_w}V_{air}^2
$
 which is also a constant ,
$
F_0=
$
initial film thickness,
$
r=X_1
$
and
$
\dot{r}=V_{sheet}=X_2
$.
\begin{figure}
    \centering
    \includegraphics[scale=1]{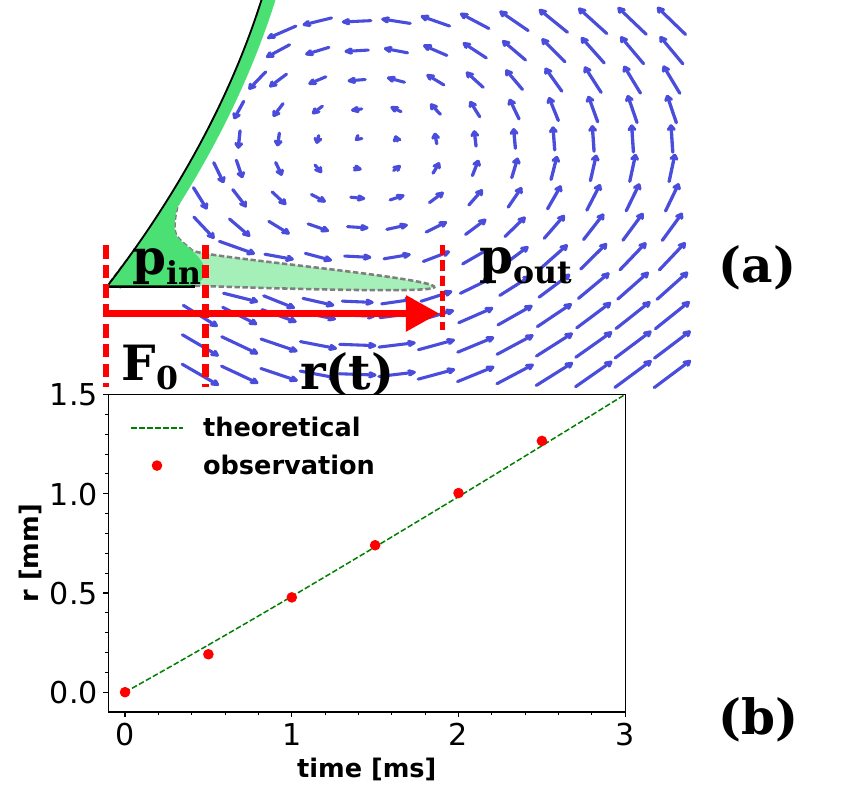}
    \caption{(a) Schematic of the sheet expansion process due to leading vortex, (b) Experimental and theoretical comparison of the sheet expansion dynamics}
    \label{Fig. 10.}
\end{figure}

%${\sigma}{\sim}0.072N/m$, 
%${\Delta}t{\sim}{2ms}=2{\times}10^{-3}s$, 
%${\rho}_{w}{\sim}10^3kg/m^3$,
%${\Delta}r{\sim}r{\sim}10^{-3}m$, 
%${\rho}_{air}{\sim}1kg/m^3$
%$$
%{{\Delta}V_{sheet}}{\sim}\left(0.144+\frac{V_{air}^2}{10^3}\right)m/s
%$$
%The approximate sheet velocity scale
%for a initial film thickness of $1mm$ is $
%{{\Delta}V_{sheet}}{\sim}0.144m/s
%$ 
Figure 10(b) shows the numerical solution of equation [8] and the experimental comparison of the sheet expansion. The constants used for the numerical solution are 
${\sigma}{\sim}0.072N/m$, 
${\Delta}t{\sim}=2{\times}10^{-6}s$, 
${\rho}_{w}{\sim}10^3kg/m^3$,
%${\Delta}r{\sim}r{\sim}10^{-3}m$, 
${\rho}_{air}{\sim}1kg/m^3$
${V_{air}{\sim}1m/s}$
The theoretical estimates corroborates well with the experimental data.
Phase E consists of a lot of exciting fluid dynamical events, as is shown in Figure 11. Fig. 12(a) represents the sheet expansion in millimeters as a function of time in milliseconds. The experimental data are marked with red dots, the experimental fit with a dashed-dotted line, and the theoretical curves predicted by the numerical solution of equation [8] by solid lines. The theoretical model predicts the sheet expansion within 1 percent accuracy. The red shaded region shows a wide variation of initial expansion velocity due to different amount of tear layer thickness. The blue shaded region represents the later sheet expansion due to the effect of the trailing jet and its corresponding air velocity field.
The initial sheet ejection leads to a more complex sheet expansion due to the trailing jet, corneal deflection, and the related capillary waves. This expansion of the sheet is shown by the blue curves. This rapid sheet expansion and acceleration cause small disturbances in the flow field to amplify and grow. The disturbances are in the azimuthal direction since each part of the sheet expands in a plane locally but in all directions. So while the sheet expands, the lighter fluid pushes the heavier fluid, and a Rayleigh Taylor instability is triggered, which amplifies small perturbations to grow into finger like structures. The disturbances that are triggered would have a predominant azimuthal symmetry if the flow field was less vigorous. However, due to other transient structures, the symmetry is broken in many places. However, the local symmetry is preserved like the distance between the fingers right ${\lambda}$ at the point of initial growth. This can be measured in a statistical sense. Fig. 11(b) shows the sheet expansion and the associated timescale as a sequence of snapshots.  Simultaneously due to the highly transient and 3D nature of the relative flow with respect to the expanding sheet, bag-like structures are formed in the expanding sheet, which undergoes breakup into daughter droplets.
   
\begin{figure}
    \centering
    \includegraphics[scale=1]{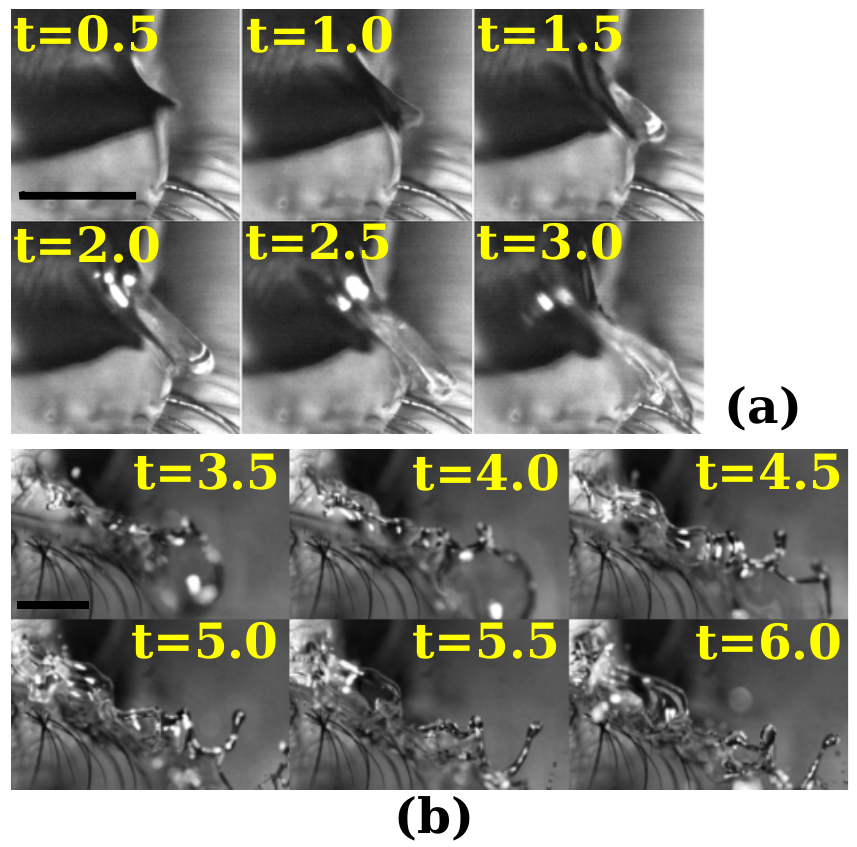}
    \caption{Time sequence of the sheet expansion (a) denotes the initial sheet ejection, (b) denotes later sheet expansion. The scale bar is 3mm.}
    \label{Fig. 11.}
\end{figure}

\begin{figure}
    \centering
    \includegraphics[scale=1]{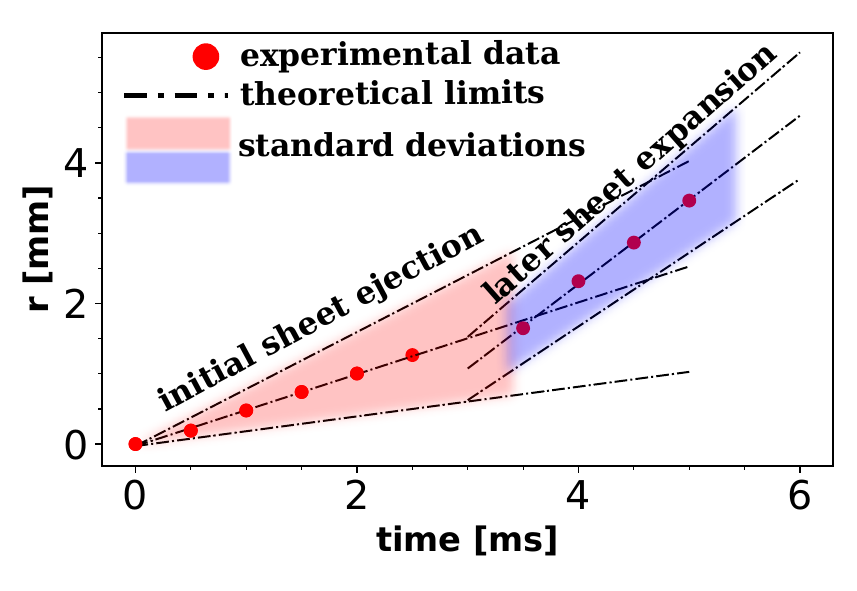}
    \caption{Sheet expansion as a function of time. Red region denotes initial sheet ejection and blue region denotes later sheet expansion.}
    \label{Fig. 12}
\end{figure}

\subsubsection*{Rayleigh Taylor Instability of the sheet}
The initial perturbations are mostly in the azimuthal direction. The wave propagation and the fluid flow is in the radial direction. The initial perturbations are due to small fluctuations in the velocity and pressure field. Therefore we can apply linear stability analysis in the azimuthal direction (1D) to get an idea of the distance between the fingers and the growth rate of the initial disturbance into the formation of fingers.  
The general Rayleigh Taylor dispersion \cite{drazin2002introduction,drazin1981hydrodynamic} relation for 1D and assuming wave propagation direction to be same as fluid flow direction which is radial in our case we have
    \begin{equation}
      \omega=\frac{1}{2}i\left(\frac{{\rho}_w-{\rho}_a}{{\rho}_w+{\rho}_a}\right)kV{\pm}(I+II-III)^{1/2}
    \end{equation}
Equation [14] depicts that the disturbance growth rate is a complex valued function. The real part of the growth rate
    has contributions from three terms $I,II,III$.
    The first term $I$ signifies the kinetic energy of one fluid with respect to the other.
    \begin{equation}
I=\frac{{\rho}_a{\rho}_w}{({\rho}_a+{\rho}_w)^2}k^2V^2
\end{equation}
The second term $II$ signifies the apparent gravity felt by the expanding sheet.
    \begin{equation}
II=\frac{({\rho}_w-{\rho}_a)}{({\rho}_w+{\rho}_a)}g_{eff}k
\end{equation}
The third term $III$ is the contribution due to the surface tension.
\begin{equation}
III=\frac{{\sigma}_{aw}}{({\rho}_w+{\rho}_a)}k^3
\end{equation}
where $V$ is the absolute value of the relative flow velocity of air with respect to the sheet and
$g_{eff}$ is the apparent gravity experienced by the sheet.

\begin{equation}
Re({\omega})=(I+II-III)^{1/2}
\end{equation}
where $Re({\omega})$ denotes the real part of the complex valued function which is physically the initial perturbations growth rate.
\begin{equation}
{Re({\omega})}^2=(I+II-III)
\end{equation}
Equation [19] shows that the kinetic energy and the apparent gravity contribute positively to the growth of the disturbance whereas surface tension has a negative effect on the growth of the disturbance and tries to suppress the growing disturbance.
$$
\frac{d}{dk}({Re({\omega})}^2)=\frac{d}{dk}(I+II-III)
$$
$$
\frac{d}{dk}({Re({\omega})}^2)=\frac{{\rho}_a{\rho}_wV^22k}{({\rho}_a+{\rho_w})^2}+\frac{({\rho}_w-{\rho}_a)g_{eff}}{{\rho}_a+{\rho}_w}-\frac{3{\sigma}_{aw}k^2}{({\rho}_a+{\rho}_w)}
$$
For extremum conditions (in this case it is a maxima) we have the derivative of the real part of the square of the growth rate with respect to the wavenumber to be zero.

\begin{equation}
\frac{d}{dk}({Re({\omega})}^2)=0
\end{equation}
\begin{equation}
\frac{{\rho}_a{\rho}_wV^22k}{({\rho}_a+{\rho_w})^2}+\frac{({\rho}_w-{\rho}_a)g_{eff}}{{\rho}_a+{\rho}_w}-\frac{3{\sigma}_{aw}k^2}{({\rho}_a+{\rho}_w)}=0
\end{equation}
The solution of equation [20] will give us the wavenumber whose real part of the growth rate is the maximum. This solution will provide the wavenumber of the fastest growing wave.
\begin{equation}
{3{\sigma}_{aw}k^2}-\frac{2{\rho}_a{\rho}_wV^2}{({\rho}_a+{\rho_w})}k-({\rho}_w-{\rho}_a)g_{eff}=0
\end{equation}

\begin{equation}
{k^2}-\frac{2{\rho}_a{\rho}_wV^2}{3({\rho}_a+{\rho_w}){\sigma}_{aw}}k-\frac{({\rho}_w-{\rho}_a)g_{eff}}{3{\sigma}_{aw}}=0
\end{equation}

\begin{equation}
k^2-Bk-C=0
\end{equation}
where $B=\frac{2{\rho}_a{\rho}_wV^2}{3({\rho}_a+{\rho_w}){\sigma}_{aw}}$
and $C=\frac{({\rho}_w-{\rho}_a)g_{eff}}{3{\sigma}_{aw}}$
\begin{equation}
k_{max}=\frac{B}{2}+\sqrt{\frac{B^2}{4}+C}
\end{equation}
Fig. 13 shows the Rayleigh-Taylor dispersion curves for three different relative velocities of the air field with respect to the sheet ($1m/s$, $4m/s$ and $8m/s$).
Fig. 14 compares the experimentally observed data of distance between fingers and the finger formation time scale as the probability density function. Fig. 14(a) depicts the probability density functions of the experimental data for the distance between the fingers, which is denoted by ${\lambda}$ in mm. The theoretical density function is generated from equation [25], and the peaks of the two distribution are very close to each other and hence corroborates the experimental and theoretical data to very high accuracy. Fig. 14(b) shows the probability density function for the finger formation timescale ${\tau}_{RT}$in ms. The Rayleigh-Taylor time scale ${\tau}_{RT}$ is the reciprocal of ${Re({\omega})}$. The green shaded region denotes the theoretical estimates of the density function, which predicts the scale correctly. The theoretical density function for ${\tau}_{RT}$ is calculated by using equation [25] in equation [18]. Equation [25] gives us the wavenumber at which the dispersion curve peaks.
\begin{figure}
    \centering
    \includegraphics[scale=1]{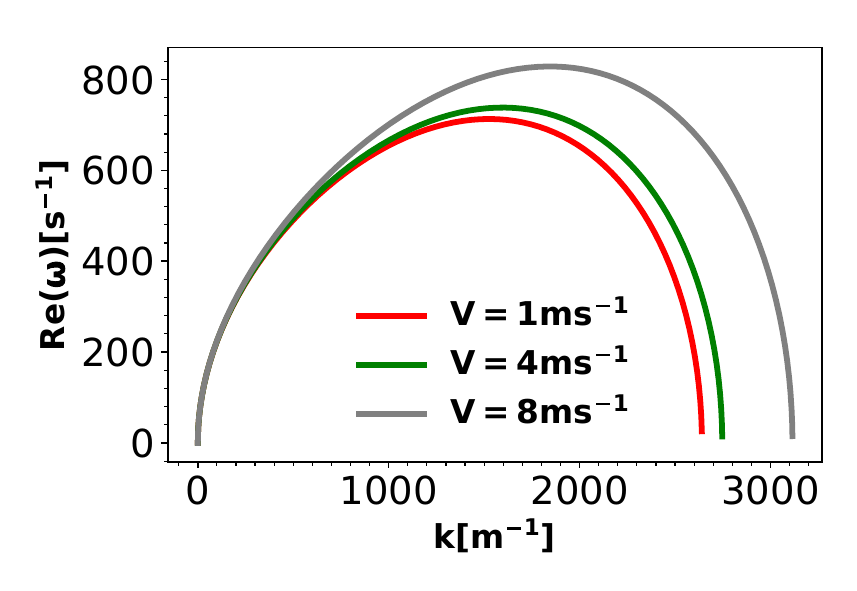}
    \caption{Rayleigh Taylor dispersion curves for different relative velocities}
    \label{Fig. 13.}
\end{figure}

\begin{figure}
    \centering
    \includegraphics[scale=1]{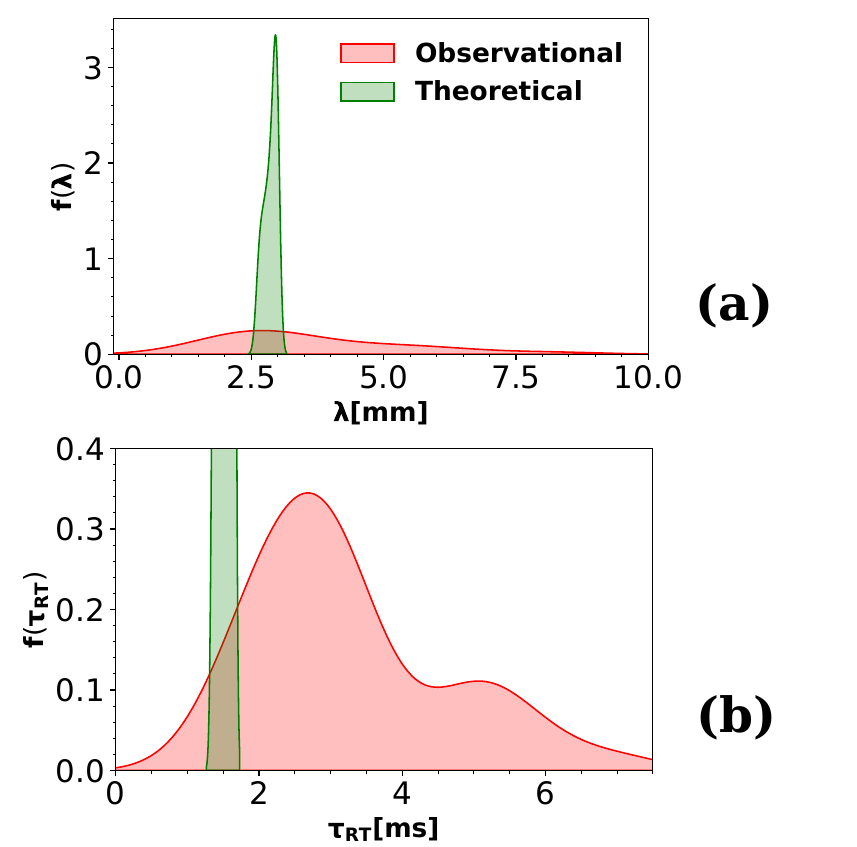}
    \caption{(a) Probability density function for the distance between the fingers.
    (b) The probability density function for the finger formation timescale.}
    \label{Fig. 14.}
\end{figure}

\subsection*{Disintegration into droplets by RP Instability (Phase F)}
The final phase of all the successive events is the disintegration of the finger-like structures into droplets by Rayleigh Plateau instability. The finger/ligament-like structures formed by the Rayleigh Taylor mechanism undergoes ligament breakup forming droplets at the tip. Fig. 15(a) depicts the ligament breakup mechanism showing how the ligament disintegrates into droplets.   
\subsection*{Rayleigh Plateau Instability}
The ligament undergoes a stretching and successively undergoes tip breakup. This stretching mechanism essentially happens in 1D locally, which approximately coincides with the sheet expansion in the radial direction. 
The dispersion relation from 1D Rayleigh Plateau stability analysis is given by \cite{drazin2002introduction,drazin1981hydrodynamic}
\begin{equation}
          \omega^2=\frac{{\sigma}_{aw}}{{\rho}_wt_0^3}kt_0\frac{I_1(kt_0)}{I_0(kt_0)}(1-k^2t_0^2)
        \end{equation}
Where $I_0$ is the  Modified Bessel function of the first kind (0th order), $I_1$ is the Modified Bessel function of the first kind (1st order), $t_0$ is the initial thickness of the ligament, $\tau_{RP}=\frac{1}{{\omega}_{max}}$ is the timescale associated with the growth rate.
$\frac{L}{t_0}>{\pi}$ is the ligament breakup criterion where $L$ is the length of the ligament, and $t_0$ is the thickness of the ligament.
All the ligament that pinches off into droplets satisfies the length to width aspect ratio condition, i.e., the aspect ratio of the ligament has to be greater than ${\pi}$ for the ligament that breakup. This is shown in Fig. 15(b). This figure shows a complete set of 60 runs. If the aspect ratio of the ligament is greater than ${\pi}$, it breakups into droplets. On the contrary, if the aspect ratio is less than ${\pi}$, the ligament would stretch without breakup. 
\begin{figure}
    \centering
    \includegraphics[scale=1]{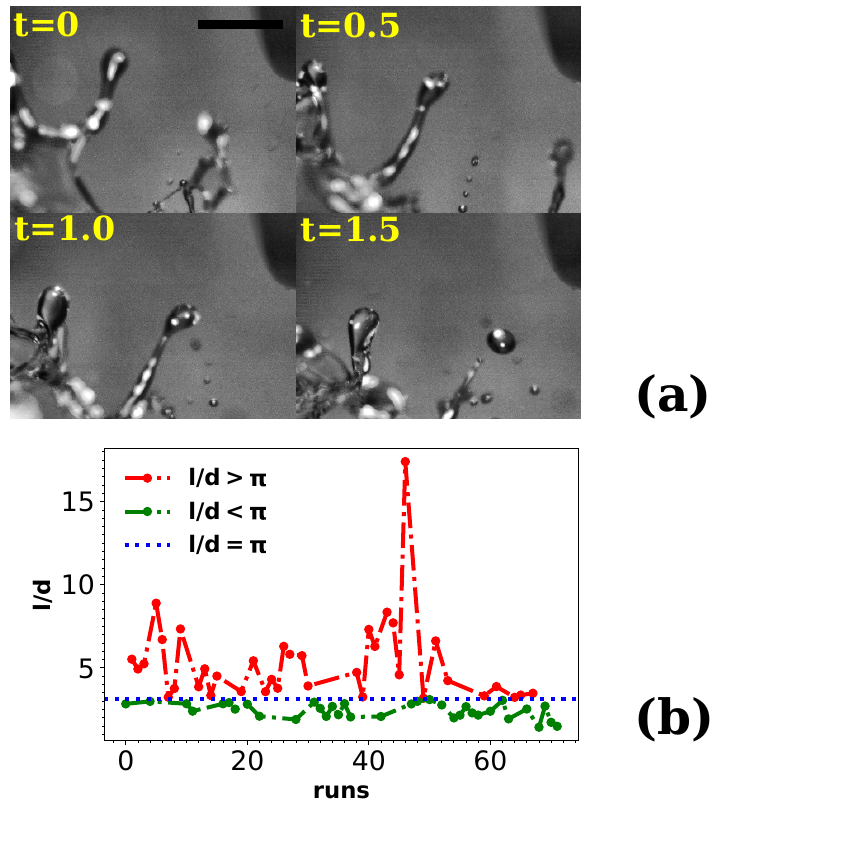}
    \caption{(a) Sequence of images showing the Rayleigh Plateau breakup and the corresponding time scale. (b) The ligament aspect ratio plotted for different experimental runs depicting the Rayleigh Plateau breakup criterion.}
    \label{Fig. 15.}
\end{figure}

Fig. 16(a) shows the theoretical dispersion curve obtained from the linear stability theory using the ligament thickness obtained from the probability density function experimentally. The distribution was experimentally obtained by assuming that the rim surrounding the expanding sheet has the same thickness as the ligament on an order of magnitude scale. This gives us an upper limit on the ligament thickness.
The black curve represents a thickness of approximately $0.22mm$, and the red curve represents a thickness of $0.43mm$. Both curves peak at the same value of wavenumber, and the corresponding wavelength is approximately $3.93mm$. However, the growth rate and the associated time scales are different. ${\tau}_{RP}$ is greater for the thicker ligament than the thinner ones. 
Fig. 16(b) depicts the experimental and theoretical probability distribution function. The theoretical and the experimental curves peak at the same value of ${\tau}_{RP}$ corroborating the linear stability analysis.
The droplets from ligament breakups have a multitude of size distribution and velocity distribution.
%The coordinate system used to characterize the size and velocity of the droplets is provided in the SI.
A bounding rectangle is used to calculate the shape of the droplet. The major axis and minor axis of the droplets were calculated from the bounding rectangle's width and height. The droplets were tracked, and the velocities were calculated from the derivatives of the position.
Fig. 17 summarizes the shape and the velocity of the droplets. Fig. 17(a) characterizes the shape of the droplets.
The droplets ejected out have a wide range of size distribution ranging from $0.2mm-3mm$. The droplets, in general, are elliptical in nature. The minor-axis to the major-axis ratio of the droplets are approximately $0.8$ on average. The droplets, on average, are elongated along the horizontal axis. The droplet velocities are also distributed over a range. The components of the droplet velocities are shown in Figure 17(b). On average, the droplets x component and y component of the velocity are approximate $0.1m/s$ approximately.  

\begin{figure}
    \centering
    \includegraphics[scale=1]{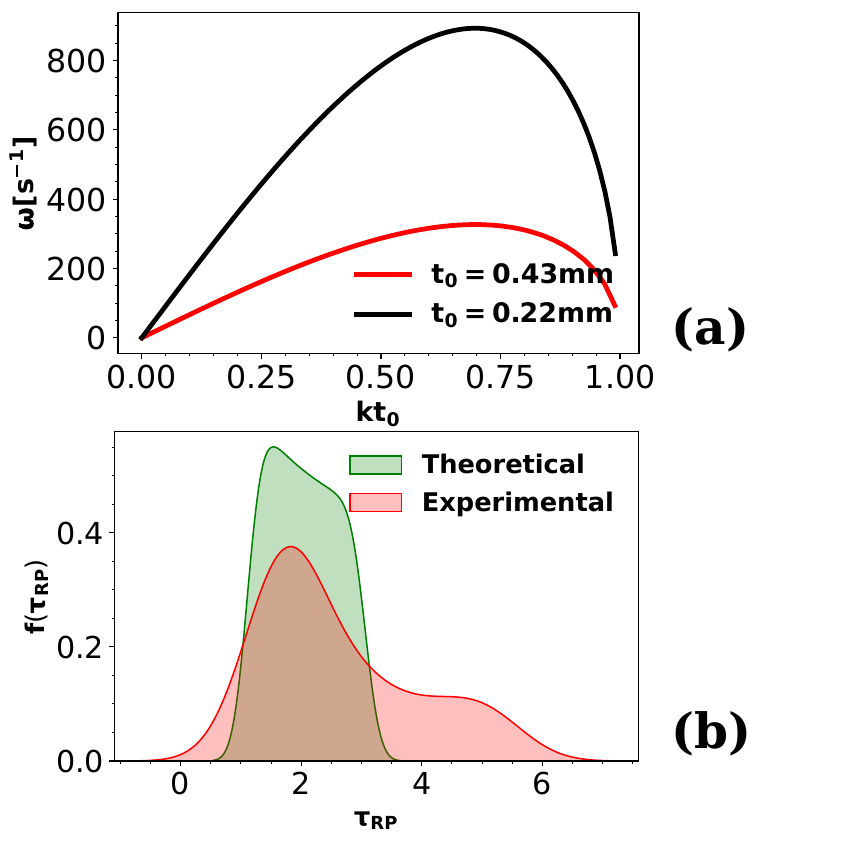}
    \caption{(a) Dispersion curve for the Rayleigh Plateau linear stability theory for different ligament thickness, (b) The probability density function for ligament breakup timescale }
    \label{Fig. 16.}
\end{figure}

\begin{figure}
    \centering
    \includegraphics[scale=1]{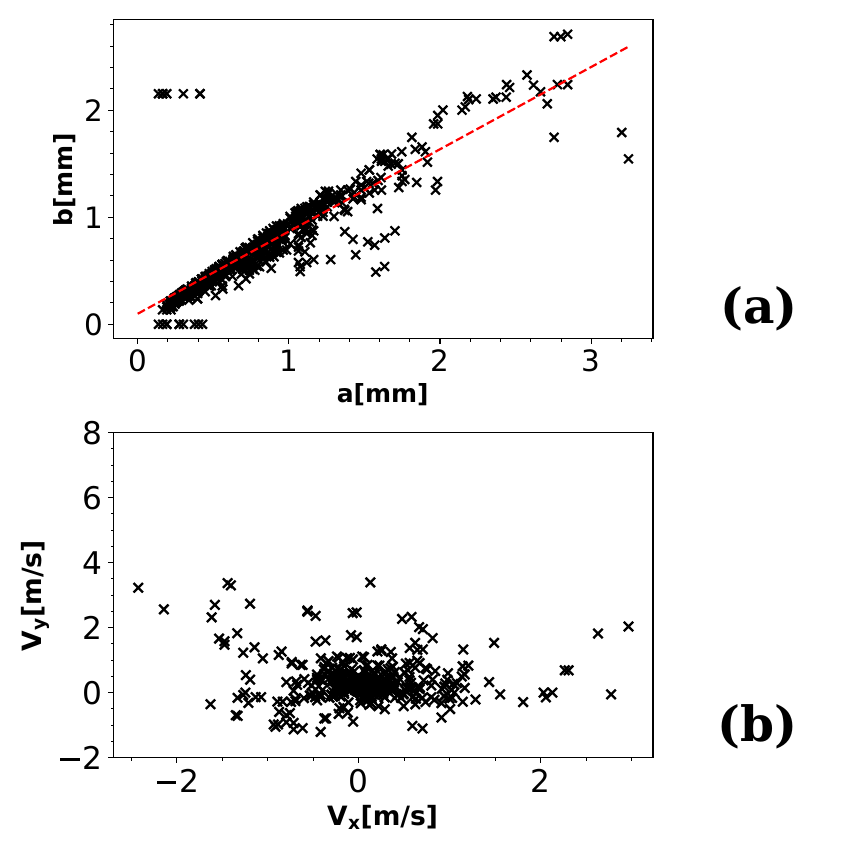}
    \caption{(a) Droplet size characterization where a is the major-axis and b is the minor-axis of the droplet respectively in mm  (b) Droplet velocity characterization showing both the x component velocity and the y component velocity.}
    \label{Fig. 17.}
\end{figure}

\section*{Conclusion}
This work addresses the fluid dynamics involved during an ophthalmologic measurement technique called tonometry used to measure the intraocular pressure (IOP) of the human eye. The mechanisms involved during tonometry were studied experimentally and theoretically in sufficient detail for dry eye and watery eye conditions using real human subjects. The experimental observations were supported by a comprehensive theoretical understanding of all the mechanisms that are involved. High-speed imaging was used to capture the intricate transient three-dimensional fluid mechanical processes that are involved. The images were acquired from different views like side, orthographic, to name a few, for a better understanding of the three-dimensional nature of the flow features. The tonometer ejects an air puff from the nozzle, which is kept at approximately $11mm$ from the eye. The air puff has been characterized using high-speed scattering and shadowgraphy techniques. It has been found that the air puff has two essential features, a leading vortex followed by a trailing jet. The leading vortex initially approaches the eye, which causes an increase in the air velocity field locally, which leads to a local pressure reduction causing an initial sheet ejection in the case of a wet eye. The wet eye is simulated by using eye drops before undergoing the tonometry measurement technique. While the sheet ejects out of the eye, the trailing jet hits the cornea causing a deflection, which in turn creates capillary waves on the surface of the eye. The sheet expands in two phases, first the ejection due to the leading vortex. Then it receives a secondary kick from the effect of the trailing jet. During the first phase of the sheet expansion, small disturbances lead to Rayleigh Taylor waves to form and grow. Due to the highly transient and three-dimensional nature of the flow field bag like structures begin to appear, it undergoes the bag breakup mode of disintegration. Further, the Rayleigh Taylor modes lead to finger like structure to form, which undergo Rayleigh Plateau break up into droplets. The droplet's size distribution and velocity were experimentally determined from the high-speed images. A complete theoretical framework relevant to the current case was developed and was compared with the corresponding experimental data.

\acknow{We would like to acknowledge Narayana Nethralaya Eye Hospital, Bangalore, India for providing us with the Non-Contact Tonometer (NCT200) for conducting the experiments.}

\showacknow{} % Display the acknowledgments section

% Bibliography
\bibliography{pnas-sample}

\end{document}